\documentclass[onecolumn]{aastex63}
\usepackage[utf8]{inputenc}
\usepackage{amsmath,amsfonts,amssymb}
\usepackage{graphicx,subfigure,multirow}
\usepackage{epsfig,color}
\usepackage{enumitem}
\usepackage{natbib}
\usepackage{hyperref}
\usepackage{verbatim}
\usepackage[bottom]{footmisc}

\begin{document}

\title{Structure of Quark Star: A Comparative Analysis of Bayesian Inference and Neural Network based modelling}



\author[0000-0002-8410-520X]{Silvia Traversi}
\affiliation{Dipartimento di Fisica e Scienze della Terra,  Università di Ferrara, Via Saragat 1, 44122 Ferrara, Italy}
\affiliation{INFN Sezione di Ferrara, Via Saragat 1, 44122 Ferrara, Italy}

\author[0000-0001-6592-6590]{Prasanta Char}
\affiliation{INFN Sezione di Ferrara, Via Saragat 1, 44122 Ferrara, Italy} 
\affiliation{Space  sciences,  Technologies  and  Astrophysics  Research  (STAR)  Institute,  Université de Liège, Bât. B5a, 4000 Liège, Belgium}

\correspondingauthor{P. Char}
\email{char@fe.infn.it}

\begin{abstract}
In this work, we compare two powerful parameter estimation methods namely Bayesian inference and Neural Network based learning to study the quark matter equation of state with constant speed of sound parameterization and the structure of the quark stars within the two-family scenario. We use the mass and radius estimations from several X-ray sources and also the mass and tidal deformability measurements from gravitational wave events to constrain the parameters of our model. The results found from the two methods are consistent. The predicted speed of sound is compatible with the conformal limit. 
\end{abstract}


\section{Introduction}
The equation of state (EOS) of dense nuclear matter is subject to extensive studies throughout the last few decades \citep{Oertel:2016bki,Baiotti:2019sew}. Still, a consensus is yet to be reached on the composition of matter at densities higher than the nuclear saturation density. In nature, such densities appear only inside the compact remnants formed after the collapse of the core of massive stars ($\gtrsim 8 M_\odot$). Therefore, observing such objects can be very useful in understanding their interior. Indeed, the increasing number of electromagnetic (EM) such as radio, X-ray and gravitational wave (GW) observations have provided valuable information on the EOS of such objects \citep{Ozel:2016oaf,Riley:2019yda,TheLIGOScientific:2017qsa,Abbott:2020uma}. The discoveries of a few pulsars over $2 M_\odot$ have put stringent constraints on the EOS of supranuclear matter \citep{Demorest:2010bx,Antoniadis:2013pzd,Fonseca:2016tux,Arzoumanian:2017puf,Cromartie:2019kug}. It requires the matter inside such compact stars (CSs) to be stiff to reach such massive stable configurations. On the other hand, the measurement of tidal deformability from the event GW170817 indicates towards smaller radii for the low mass CSs \citep{Abbott:2018wiz}, meaning the EOS to be soft at the densities corresponding to the low mass configurations. 

From the perspective of nuclear physics, the theory of nuclear interaction at such densities is not fully known. Although, it is expected that new strange degrees of freedom should appear in the system as the the core densities increase, which in turn would soften the EOS reducing the maximum mas below the observed pulsar masses \citep{Glendenning}. The standard way to approach this problem is to introduce repulsive interaction to make the EOS stiffer \citep{Chatterjee:2015pua}. But, this exercise would also make the low mass stars larger contradicting the GW data. Additionally, quark stars (QSs) entirely made of strange quark matter (SQM) can also exist if the Bodmer-Witten hypothesis of the SQM to be the most stable state of matter holds true \citep{Bodmer:1971we,Witten:1984rs}. But, all the CSs can not be QSs as the pulsar glitches are not possible to explain without assuming a star with a crust \citep{Alpar:1987vk}. These observational contradictions with the theoretical understanding of dense matter physics led to the proposition of an alternative scenario namely "two-families scenario" where disjointed families of QSs and hadronic stars (HSs) can coexist \citep{Drago:2013fsa,Drago:2014oja,Drago:2015cea}. In this model, the smaller stars are HSs composed of several strange baryons and $\Delta$ resonances and the massive stars are the QSs with deconfined quark matter. There exist also other models in literature concerning two disconnected branches of compact objects. The most popular being the so called "twin-stars" solution where the most massive stars are hybrid stars with a sharp phase transition. This branch has a characteristic feature of smaller radii, as opposed to the "two-families scenario". In recent years, many analyses have been performed to test the "twin-stars" model against the observational data \citep{Montana:2018bkb,Christian:2019qer,Blaschke:2020qqj} and a comparison with the "two-families scenario" has also been provided in \cite{Burgio:2018yix}.

The connection between nuclear physics and astrophysical observations is usually reflected in the one-to-one correspondence between the EOS and the mass-radius ($M-R$) relations, calculated from the Tolman-Oppenheimer-Volkoff (TOV) equations \citep{TOV,Lindblom:1998dp}. In recent times, tidal deformability ($\Lambda$) has also been used as a complimentary information to the radius measurements \citep{Hinderer:2007mb, Damour:2009vw,Hinderer:2009ca}. Therefore, one can also map the EOS to the $M-\Lambda$ relations. The usual strategy to estimate the EOS is to build a certain model consisting a number of parameters. One can have some knowledge of these parameters a-priori from physical considerations. Then, one can systematically update that knowledge with the observational data using Bayesian inference methods into the posterior distributions of those parameters \citep{Steiner:2010fz}. This is a robust statistical method where one can quantify the feasibility among competing EOS models, the prior dependence on the inference. In reality, the scope of this method is limited by the existence of only a few observations and also the observational points are not distributed optimally throughout the $M-R$ plane to probe the whole plausible range of the EOS inside a CS. It is therefore needed to develop a methodology as a complimentary approach to the standard parameter estimation method. A machine learning based prediction method can be used as an alternative procedure whose application to high energy physics, astrophysical data analysis and other branches of physics has gained momentum of late. Deep learning techniques using neural networks (NN) has also been used specifically to estimate the dense matter EOS \citep{Fujimoto:2017cdo,Fujimoto:2019hxv,Morawski:2020izm}. In this studies, a particular parameterization of the EOS, namely piecewise polytropes has been used to train, validate and test the NN. In another work, two different learning methods, support vector machine regression and deep learning utilizing NN, have been compared to study the saturation properties of nuclear matter in terms of nuclear empirical parameters \citep{Ferreira:2019bny}. 

The purpose of the present work is twofold. Firstly, we wish to present a comparative analysis between the Bayesian parameter estimation and NN based prediction. While doing that we also investigate the structure and properties of QSs within the two-families scenario as a followup study to our previous work on HSs. The paper is organized as follows. In section \ref{twofamilies}, we describe the EOS model for QSs and the calculation of its structure. The sources used and their selection criteria are specified in section \ref{sources}.  In section \ref{bayes} and \ref{DLNN}, we explain our methodology. Finally, in section \ref{results} and \ref{summary}, we discuss our results and summarize.

\section{Two family Scenario and Quark Star} \label{twofamilies}
At the core, the two-families scenario utilizes the idea of the absolute stability of the strange quark matter. One can also present several arguments based on astrophysical observations to make a case for this idea as explained in \cite{Drago:2013fsa,Drago:2014oja}. In this scenario, it is possible to get very compact stars as hadronic stars with radius smaller than $12$ km as well as very massive stars as quark stars with maximum mass about $\sim 2.2 M_\odot$. In a previous work, we have explored the parameter space of relativistic mean field model to build compact hadronic stars \citep{Traversi:2020aaa}. In this work, we mainly focus on the quark stars (QS). For simplicity, we take a constant-speed-of-sound EOS for the QSs \citep{Alford:2013aca,Zdunik:2012dj,Chamel:2012ea,Drago:2019tbs} in which the relations between the pressure, energy density and baryon density are as follows,

\begin{eqnarray}
 p &=& c_{s}^{2} \left(e-e_0\right), \\
 p &=& \frac{c_{s}^{2} e_0}{c_{s}^{2} + 1} \left( \left( \frac{n}{n_0} \right)^{c_{s}^{2} + 1} - 1\right).
\end{eqnarray}
 Here, $e_0$ and $n_0$ represent the energy density and the baryon density at zero pressure, respectively. There are three main quantities in this parameterization: the speed of sound ($c_s$), the $n_0$ and the energy per baryon $(E/A)_0 = e_0 /n_0$. The bounds on the energy per baryon come from the stability of iron nuclei in light of the two-flavor and three-flavor quark matter. The condition for the absolute stability of the three-flavour strange quark matter is $(E/A)_0 < 930$ MeV whereas to keep the Fe$^{56}$ stable against decaying in two-flavor quark matter $(E/A)_0 > 830$ MeV \citep{Weissenborn:2011qu,Drago:2019tbs}. Usually, the speed of sound in the hadronic matter is less than $\sqrt{1/3}$ as the EOS is soft, while at high densities in quark matter , theoretical calculations suggest that it should reach the conformal limit of $\sqrt{1/3}$, due to the QCD asympotic freedom \citep{Bedaque:2014sqa}. The validity of this limit is presently an object of discussion and it has been tested against the recent observational data of very massive pulsars and the limits on the tidal deformability provided by the GW events \citep{Reed:2019ezm,Annala:2019puf,Marczenko:2020wlc}. We will further investigate this subject in the context of the "two-families scenario" in a follow-up paper currently in preparation.

 \subsection{Calculation of the Structure}
 The configuration of a static and spherically symmetric CS is modelled using the TOV equations of hydrostatic equilibrium as,
 
 \begin{eqnarray}
  \frac{dp}{dr} &=& -\left(e+p\right)\frac{m+4\pi r^3 p}{r\left(r-2m\right)} \label{eq:dpdr}\\
  \frac{dm}{dr} &=& 4\pi r^2 e. \label{eq:dmdr}
 \end{eqnarray}

Here, $m=m(r)$ is the enclosed gravitational mass at radius $r$ from the center. We integrate \ref{eq:dpdr} and \ref{eq:dmdr} from the center $r=0$ to the surface of star at radius $R$ where $p(R)=0$, leading to its mass $M=m(R)$. Additionally, we also calculate the tidal deformability of the star, defined as

\begin{equation}
    \Lambda = \frac{2}{3} k_2 \left(\frac{R}{M}\right)^5,
\end{equation}
where $k2$ is the electric-type tidal Love number associated with the quadrupolar tidal perturbation, given by the following expression

\begin{eqnarray} 
k_2 &=& \frac{8C^5}{5} (1-2C)^2[2+2C(y-1)-y] \nonumber\\
&\times&\Big\{2C[6-3y+3C(5y-8)]+4C^3[13-11y+C(3y-2)+2C^2(1+y)] \nonumber\\
&+& 3(1-2C)^2[2-y+2C(y-1)]\ln{(1-2C)}\Big\}^{-1}.
\end{eqnarray}
Here, $y$ is the solution of the following equation at $r=R$,

 \begin{equation}
 \frac{dy}{dr} = -\frac{y^2}{r} - \frac{r+4\pi r^3 (p-e)}{r(r-2m)}y + \frac{4(m+4\pi r^3 p)^2}{r(r-2m)^2} + \frac{6}{r-2m} 
 -\frac{4\pi r^2}{r-2m}\left[5e+9p
 +\frac{e+p}{(dp/de)}\right] .
 \end{equation}

Since, the QSs have a sharp discontinuity of energy density at the surface, the value of $y(R)$ requires a correction term \citep{Hinderer:2009ca,Postnikov:2010yn,Takatsy:2020bnx} as,

\begin{equation}
 y = y(R) - \frac{4\pi R^3 e_{-}}{M}.
\end{equation}

Here, $e_{-}$ is the energy density just inside the surface.

\section{Observational Data: Candidate Quark Stars}\label{sources}
In this section, we specify the sources used in this work. We select some of the sources form the X-ray measurements from \cite{Ozel:2015fia} depending on their masses \footnote{The M-R distributions of the sources of \cite{Ozel:2015fia} are available at \href{http://xtreme.as.arizona.edu/neutronstars/}{http://xtreme.as.arizona.edu/neutronstars/}.}.  \\
In the context of the two families scenario, three types of binaries are possible: HS-HS, HS-QS, QS-QS. An estimate of the threshold mass for a prompt collapse to black hole, $M_{thr}$, has been provided in \cite{DePietri:2019khb} for each of the different combinations: in the case of HS-HS we found $M_{thr}=2.5 M_{\odot}$.
Since GW170817 was not a prompt collapse event, we interpret it as a HS-QS merger and thus we classify the high-mass component as a QS and label it as GW170817\_1. Then, we marginalize the distribution of the mass of the object from the posterior samples and find the mean value as $1.49 M_\odot$.
Therefore, the sources with the mean of the mass distribution, $M \gtrsim M_{\mathrm{GW170817\_1}}$, can be hypothetically identified as QSs. 
Explicitly, we have considered 4U 1724-07, SAX J1748.9 2021, 4U 1820--30,  4U 1702--429,  J0437--4715,  GW170817\_1, GW190425\_1 and GW190425\_2. 
For 4U 1702--429 \citep{Nattila:2017wtj} and J0437--4715 \citep{Gonzalez-Caniulef:2019wzi}, we take the following form of a bivariate Gaussian distribution to mimic the $M-R$ posterior since the full distribution is not available,

\begin{equation}
    P(M, R) = \frac{1}{2\pi \sigma_M \sigma_R \sqrt{1-\rho^2}} \exp{\{ -\frac{1}{2(1-\rho^2)} [\frac{(M-\mu_M)^2}{\sigma_M^2}-2\rho \frac{(M-\mu_M)(R-\mu_R)}{\sigma_M \sigma_R} +\frac{(R-\mu_R)^2}{\sigma_R^2}]\}}, 
    \label{eq:mr_post}
\end{equation}

For 4U 1702--429, we use $\mu_M = 1.9 \rm M_{\odot}$, $\mu_R = 12.4 \rm km$, $\sigma_M = 0.3 \rm M_{\odot}$, $\sigma_R = 0.4 \rm km$, and $\rho = 0.9$, as before to represent the correlation between the measurements, since these were simultaneous measurements. For J0437--4715, we use $\mu_M = 1.44 \rm M_{\odot}$, $\mu_R = 13.6 \rm km$, $\sigma_M = 0.07 \rm M_{\odot}$, $\sigma_R = 0.85 \rm km$, and $\rho = 0.0$, since the mass and radius measurements were independent in this case. We have chosen this particular source despite its mass being lower than $M_{\mathrm{GW170817\_1}}$, because it has a radius larger than $\sim 13$ km and HSs in our scheme can not produce such large radius. Therefore, it is assumed to be a quark star. 
For the GW sources, we take directly the distribution for their individual $\Lambda$s, as converting to $M-R$ posterior requires assumption of certain universal relations which do not include quarks and postulate of all CSs having identical EOS \footnote{The data from GW170817 and GW190425 are available at \href{https://dcc.ligo.org/LIGO-P1800115/public}{https://dcc.ligo.org/LIGO-P1800115/public} and, 
\href{https://dcc.ligo.org/LIGO-P2000026/public}{https://dcc.ligo.org/LIGO-P2000026/public}}. 


\section{Bayesian Inference}\label{bayes}
We use the Bayesian framework developed in  \cite{Steiner:2010fz,Ozel:2015fia, Raithel:2017ity}. Bayes' theorem tells us that the posterior distribution function (PDF) of a set of parameters ($\theta_j$) given a data ($D$) for a model ($M$) can be expressed as, 
\begin{equation} \label{E:posterior}
    P(\theta_j|D,M) = \frac{P(D|\theta_j,M)P(\theta_j|M)}{P(D|M)},
\end{equation}
where $P(\theta_j|M)$ is the prior probability of the parameter set $\{\theta_j= e_0, c_s^2\}$, $P(D|\theta_j,M)$  is the likelihood function of the data given the model, and $P(D|M)$ is known as evidence for the model. For a given data set $P(D|M)$ is a constant and can be treated as a normalization factor. Hence, we have have in this case,

\begin{equation}\label{E:poste_para}
    P(e_0, c_s^2|\textrm{data}) = C P(\textrm{data}|e_0, c_s^2) P(e_0) P(c_s^2),
\end{equation}

where,  $P(e_0)$, $P(c_s^2)$ are the priors over $e_0$ and $c_s^2$; and

\begin{equation}
     P(\mathrm{data}|e_0, c_s^2) = \prod_{i=1}^{N} P_i(M_i,R_i|e_0, c_s^2) \qquad \textrm{or,} \qquad
     P(\mathrm{data}|e_0, c_s^2) = \prod_{i=1}^{N} P_i(M_i,\Lambda_i|e_0, c_s^2)
\end{equation}

is the likelihood of generating $N$ observations given a particular set of EOS parameters. We follow the procedure suggested by \cite{Raithel:2017ity} to calculate the probability of the realization of $(M, R)$ or $M,\Lambda$ for a particular source given an EOS. We take a set of parameters to construct the EOS, solve the TOV equations and build a $M-R-\Lambda$ sequence up to the maximum mass which corresponds to the last stable point of the curve. After that, we compute the probability of each configuration of the curve using the $M-R$ or $M-\Lambda$ distribution of the source. Finally, we assign to the parameter set the maximum probability obtained for the configurations as,

\begin{equation}
    P_i(M_i,R_i|) = P_{\textrm{max}}(M_i,R_i|e_0, c_s^2, e_c)  \qquad \textrm{or,} \qquad P_i(M_i,\Lambda_i|) = P_{\textrm{max}}(M_i,\Lambda_i|e_0, c_s^2, e_c),
\end{equation}

where, the $M-R-\Lambda$ sequence for a given EOS is parameterized by the central energy density ($e_c$) of the star.  We use the Markov-Chain Monte Carlo (MCMC) simulations to populate the posterior distribution of equation (\ref{E:poste_para}) using the python based software \textsf{emcee} with stretch-move algorithm \citep{ForemanMackey:2012ig}.

\section{Deep Learning}\label{DLNN}
Deep learning method is used to create complex nonlinear mapping between the input and output. It relies on the NN optimized by a set of training data to be able to predict the most likely output given an input. The advantage of NN to predict the EOS is that the multilayered structure of the NN is capable of reproducing the nonlinear nature of the inversion mapping between the $M-R$ relation and EOS, excluding the uncertainties about the assumption of a fitting function. In this work, we mainly follow the methodology developed in  \cite{Fujimoto:2017cdo,Fujimoto:2019hxv} in constructing the NN and preparing the data. The model function of the NN can be written as:

\begin{equation}
    y^{(k+1)}_{i} = f^{(k+1)} \left( \sum^{N_{k}}_{j=1} W^{(k+1)}_{ij} y^{(k)}_{j} + a^{(k+1)}_{i} \right).
\end{equation}

Here we have, the $k$-th layer containing $N_{k}$ numbers of neurons, $y^{(k)}_{i}$ are the values stored at the neurons at the $k$-th layer, $\{W^{(k)}_{ij}, a^{(k)}_{i}\}$ being the weights and biases respectively on the $k$-th layer, and finally, the $f^{(k)}$ being the so-called activation functions. These transformations are applied iteratively between subsequent layers. For the optimization procedure, a loss function can be chosen and thus minimized.

\subsection{Preparation of Training, Validation and Test Data}
We follow the similar procedure described in \cite{Fujimoto:2017cdo,Fujimoto:2019hxv} to generate the training and validation data set. We select randomly a number $(N_{\mathrm{EOS}})$ of parameter sets for which we construct the EOSs and calculate the $M-R-\Lambda$ sequences. Since our quark stars should be massive, we keep in the sequence only the stars with masses within $[1.3 M_\odot, M_{\mathrm{max}}]$. Next, we sample total 8 data points, representing the sources described in \ref{sources}, from a uniform distribution of $M$ over the remainder of the sequence. We now have 5 points of $(M^0_i, R^0_i)$ and 3 points of $(M^0_i, \Lambda^0_i)$. Now, for the network to learn the observational errors associated with the sources and connect with the "true" sequence, one has to introduce certain shifts on the bare value of the parameters. Since we already have a set of $(\sigma_M, \sigma_R)$ and $(\sigma_M, \sigma_\Lambda)$ after marginalization of the observational data, we build Gaussian distributions for the mass with mean at $M_i^0$ and standard deviation $\sigma_M^i$, for the radius with mean at $R_i^0$ and standard deviation $\sigma_R^i$ and for the tidal deformability with mean at $\Lambda_i^0$ and standard deviation $\sigma_\Lambda^i$. From these distributions, we sample 5 new pairs of $(M_i, R_i)$ and 3 pairs of $(M_i, \Lambda_i)$. In this way, we take into account the observational errors of the real data. Now, for each of the selected EOS, we repeat the last step a large number of times $(N_s)$ and finally produce $N_{\mathrm{EOS}} \times N_s$ numbers of training data. Each of these data point is a vector of 16 entries that comprises of the masses, radii and tidal deformabilities $(M_i,M_j,R_i,\Lambda_j; i=1,2,..,5; j=6,7,8)$ . For the validation and the test set, we repeat the same exercise but with smaller $N_{\mathrm{EOS}}$ and $N_s = 1$. The final size of our training data set is $2000 \times 100$ in the case of $1D$ and $10000 \times 100$ for $2D$ while the validation and test set in both cases are $1000 \times 1$. 
Once the network is trained, it should be able to give the value of the parameters of the "real" EOS starting from the mean values of the marginalized distributions of the masses and radii/tidal deformabilities of the selected sources.

\subsection{Features of the Neural Network}
Here, we specify the details of the NN used in this calculation. We use the Python package Keras \citep{chollet2015keras} using TensorFlow \citep{abadi2016tensorflow} as a backend and Scikit-learn \citep{pedregosa2011scikit}. The structure of our network is is summarized in Table \ref{tab:NN}.

\begin{table}[h!]
    \centering
    \begin{tabular}{|c|c|c|}
    \hline
    \hline 
   
    Layer index & nodes & activation \\
    \hline 
    0 & 16 & N/A\\
    \cline{1-3}
    1 & 50 & ReLu\\
     \cline{1-3}
    2 & 50 & ReLu  \\
    \cline{1-3}
    3 & 50 & ReLu  \\
    \cline{1-3}
    4 & 1/2 & tanh \\
    \hline 
    \end{tabular}
    \caption{Construction of the NN in this study. 16 neurons at the input layer correspond to 5 pairs of mass and radius and 3 pairs of mass and tidal deformability.}
    \label{tab:NN}
\end{table}

So, the input layer contains the same number of neurons as the observed parameters of the sources (16, in this case). The output layer contains the number of neurons as the number of the EOS parameters (1 or 2 depending on our choice of QS EOS). For the training, validation and test data, we use the 'StandardsScaler' function of Scikit-learn to normalize the observational part and the EOS part is normalized with uniform normalization $(y_{norm} = \frac{y-y_{min}}{y_{max}-y_{min}})$.

We select the activation function 'tanh' for the output layer to get a normalized output which can be converted to our desired parameter values afterwards. For the internal layers, we use the standard 'ReLU' function.
We choose mean squared logarithmic error (MSLE) as the loss function. 

\begin{equation}
        MSLE = \frac{1}{m} \sum_{i=1}^{m} {(\ln{(y^i +1)} - \ln{(y_p^i +1)})}^2,
\end{equation}
where, $y$ are the labels i.e. the real values of the parameters and $y_p$ are the predicted values.
We use the standards 'Adam' optimization method \citep{Kingma:2014vow} along with mini-batch size 128. Finally, the NN parameters are initialized with Glorot uniform distribution \citep{pmlr-v9-glorot10a}. For the $2$-parameter EOS calculation, the '$l2$' regularization is used. The learning rate is taken as $\alpha = 0.005$.
Due to the limitation of the computational resources, we have not performed the hyperparameter tuning which we have set aside as a future exercise. 

The uncertainties in the NN prediction can be estimated using the root-mean-squared error ($\mathrm{RMSE} = \sqrt{\sum_{i=1}^m \frac{(y^i - y_p^i)^2}{m}}$) estimated on the test set. For the $2D$ case, we find RMSEs separately for each of the parameters then we built a $2D$ Gaussian using the RMSEs as sigmas. Then, we take the 68\% CI from that distribution and calculate the corresponding quantities.

\section{Results and discussions} \label{results}
\begin{table}[b!]
    \centering
    \begin{tabular}{|c|c|c|c|c|}
    \hline
    \hline 
   
     & Bayes 1P & NN 1P & Bayes 2P & NN 2P \\
    \hline 
    $e_0$ (MeV fm$^{-3}$) & $191.84$ & $191.04$ & $183.48$ & $191.29$ \\
    \cline{1-5}
    $c_s^2$  & 1/3 & 1/3 & $0.306$ & $0.38$ \\
    \cline{1-5}
    $M_{max} (M_{\odot}$) & $2.18$ & $2.19$ & $2.13$ & $2.37$ \\
    \cline{1-5}
    $R_{M_{max}}$ (km)  & $12.01$ & $12.03$ & $12.00$ & $12.42$ \\
    \cline{1-5}
    $R_{1.6}$ (km) & $12.10$ & $12.12$ & $12.20$ & $12.26$ \\
     \cline{1-5}
    $\Lambda_{1.6}$ & $368$ & $373$ & $382$ & $417$\\
    \hline 
    \end{tabular}
    \caption{Features of the most probable EOS obtained with the Bayesian analysis (Bayes) and the NN method (NN) in the one parameter (1P) and two parameter (2P) case. The inferred values are specified together with the corresponding maximum mass ($M_{max}$) star and its radius ($R_{M_{max}}$) and for the $1.6 M_{\odot}$ configuration, the radius ($R_{1.6}$) and the tidal deformability ($\Lambda_{1.6}$).}
    \label{tab:max_values}
\end{table}
We present our results for Bayesian and NN calculations following the methodology developed in \ref{bayes} and \ref{DLNN} respectively for two cases in separate subsections, first keeping the $c_s^2$ fixed, only varying the $e_0$, and then varying both of them. The goal is to understand which value of the squared speed of sound is the most suitable to describe the observational data within our model. Then, we check whether or not our result is able to fulfill the conformal limit suggested by QCD calculations.

As the most massive pulsar discovered till now has a mass of $2.14_{-0.09}^{+0.10}$ \citep{Cromartie:2019kug}, we put a strict lower limit of $2.05 M_\odot$ for the maximum mass in our calculation. We use all the sources listed in \ref{sources}. 
Table \ref{tab:max_values} summarizes our results concerning the most probable EOSs. We use the python \textsf{corner.py} package to visualize one- and two-dimensional projection plots of the samples \citep{corner}. 

\subsection{One parameter case: $e_0$}\label{sec:e0}
\begin{figure}[t!]
    \centering
       \includegraphics[width=0.25\textwidth]{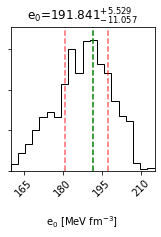}
       \caption{PDF for the parameter $e_0$ from the Bayesian analysis. The green line is placed on the mode of the distribution, while the red line represent the 68\% CI.}
    \label{fig:e0}
\end{figure}
\begin{figure}[b!]
    \centering
        \begin{tabular}{cc}
         \includegraphics[width=0.5\textwidth]{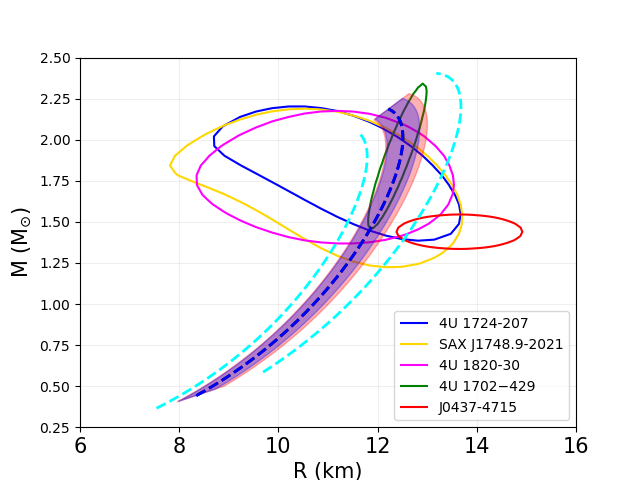}& \includegraphics[width=0.45\textwidth]{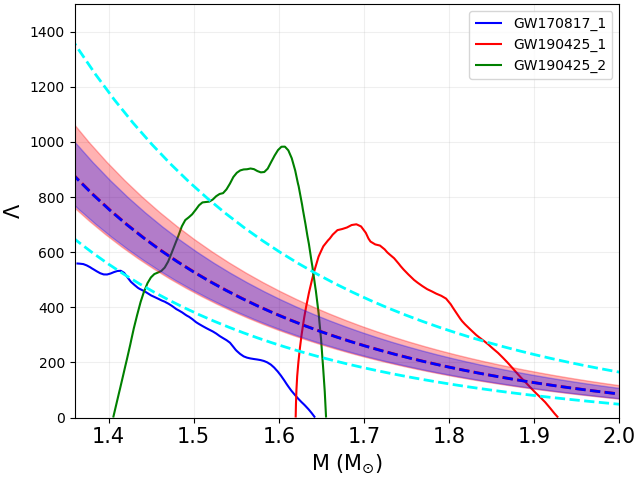}
         \end{tabular}
        \caption{Comparison between the M-R (left) and M-$\Lambda$ curves obtained with the two methods employed. The red and blue shaded regions correspond to the 68\% CI predicted by the Bayesian analysis and NN respectively. The two most probable configurations, which are indistinguishable, are plotted as the dashed blue line. Finally, the dashed cyan lines represents the border of the allowed parameter space.}
    \label{curves_1p_comparison}
\end{figure}
This is the simplest case in our analysis with only one parameter. The prior range for $e_0$ is between $160$ and $232$ MeV fm$^{-3}$, which corresponds to the allowed interval in $(E/A)_0$ for the Witten hypothesis to hold true. In this case, we have fixed $c_s^2 = 1/3$. Values of $e_0$ larger than about $220$ MeV fm$^{-3}$, corresponding to the softest EOSs, are automatically ruled out by the maximum mass limit. In Fig. \ref{fig:e0}, we have shown the most probable value of $e_0$ along with the $1\sigma$ error from the Bayesian calculation. The distribution peaks at $191.841$ MeV fm$^{-3}$. The maximum mass for the sequence corresponding to most probable value of $e_0$ is $2.18 M_\odot$ and the radius of the $1.6 M_{\odot}$ configuration is $R_{1.6}=12.10$ km.\\
From the trained NN, we get the predicted value for $e_0$ as $191.04$ MeV fm$^{-3}$, almost equivalent to the previous one, with the error $\mathrm{RMSE} = 10.03$ MeV fm$^{-3}$ estimated on the test set. The corresponding $M_{max}$ is $2.19 M_\odot$ and the $R_{1.6}=12.12$ km. In Fig. \ref{curves_1p_comparison}, we compare the $M-R$ and $M-\Lambda$ plots with the values from the $68\%$ confidence interval (CI) of the posterior distribution from the Bayesian calculation (shaded red) and NN predicted (shaded blue) range. The dashed cyan line corresponds to the initial range of the parameter used in both calculations. The most probable configuration from the Bayesian inference coincides with predicted value from the NN (dashed blue) and the errors are also quite similar.  Also in both cases the maximum masses achieved from the most probable parameters are consistent with presently accepted values \citep{Rezzolla:2017aly}. The result is exactly what we expect while comparing both methods. \\
At 68\% level, these results are not in agreement with the source J0437--4715 and also the GW170817\_1. While GW170817\_1 indicates towards a smaller radius, J0437--4715 directs towards a bigger radius. 


\subsection{Two parameter case: $e_0$ and $c_{s}^{2}$}\label{sec:e0cs}
\begin{figure}
    \centering
     \includegraphics[width=0.5\textwidth]{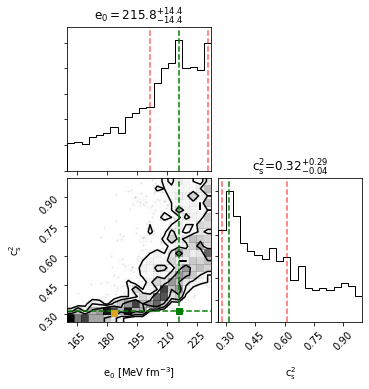}
    \caption{Joint PDF from the Bayesian analysis for the parameters $e_0$ and $c_s^2$. In addition, the marginalized distributions where the green lines are placed on the modes, while the red line represent the 68\% CI. The yellow point indicates the maximum of the 2D posterior.}
    \label{fig:e0_cs}
\end{figure}

\begin{figure}
    \centering
        \begin{tabular}{cc}
         \includegraphics[width=0.5\textwidth]{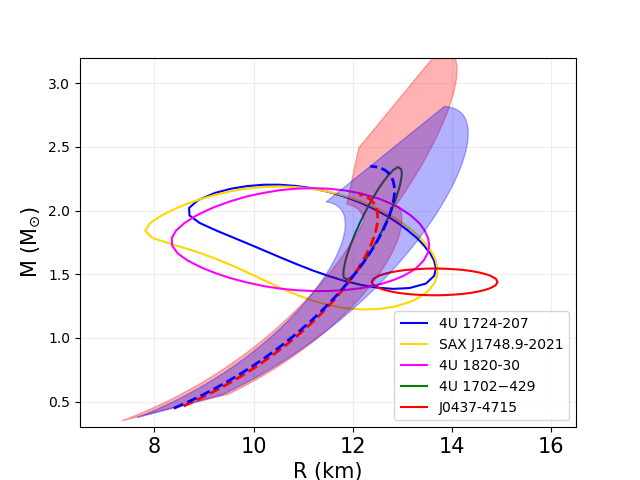}& \includegraphics[width=0.45\textwidth]{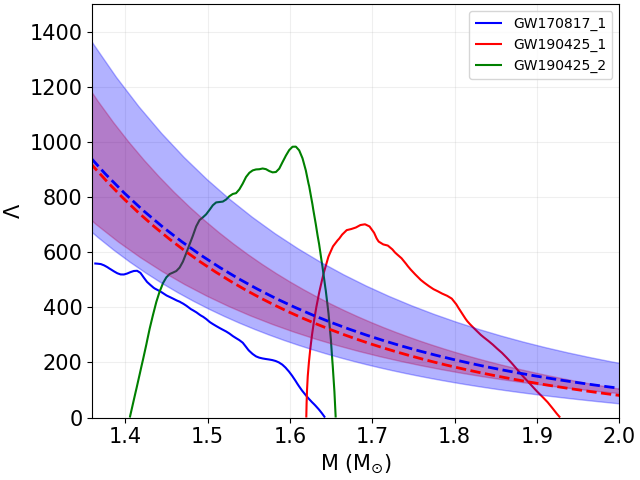}
         \end{tabular}
        \caption{Comparison between the M-R (left) and M-$\Lambda$ curves obtained with the two methods employed. The red and blue shaded regions correspond to the 68\% CI predicted by the Bayesian analysis and NN respectively. The two most probable configurations are plotted as the dashed red (Bayes) and blue (NN) lines.}
    \label{curves_2p_comparison}
\end{figure}
Next we present the case where we characterize the speed of sound as a free parameter and try to find out how far its value can deviate from the conformal limit given the present data. We build a joint posterior for $e_0$ and $c_s^2$ given the priors on $e_0$ between $160$ and $232$ MeV fm$^{-3}$, same as before, and for $c_s^2$, the range [0.1,1]. In this case, the interval in $e_0$ is not reduced a priori by means of the maximum mass limit, since an increase of the sound speed allows also to configurations with $e_0>220$ MeV fm$^{-3}$ to reach masses larger then $2.05 M_{\odot}$. On the contrary, the maximum mass constraint imposes a lower bound for $c_s^2$ at about $0.26$.    
In figure \ref{fig:e0_cs}, we present the marginalized PDFs for $e_0$ and $c_s^2$ along with the most probable values and $1\sigma$ errors, while in the $2D$ projection plot $1\sigma (39.3\%)$, $68\%$ and $90\%$ CI are shown.  The posterior reveals a correlation among the parameters. We find two classes of solutions with low and high values of ($e_0$, $c_s^2$). The second class of values are inside the 68\% CI albeit individually less probable from the low ($e_0$, $c_s^2$) points.
For this reason, the marginalized distributions of $e_0$ peaks at a value which is quite distant from the high probability region of the $2D$ PDF. 
The most probable point of the joint PDF is at $e_0=183.48$ MeV fm$^{-3}$, $c_s^2 = 0.306$, but we also find  other points with very similar probabilities close to the maximum. These two different classes are also found in the $M-R$ sequences in figure \ref{curves_2p_comparison}. 
The first one is characterized by not too large maximum masses $ \sim 2.1-2.2 M_{\odot}$ and radii in the range $R_{1.6}\sim 11.6-12.9$ km allowing the 68\% CI to overlap also with J0437–4715. On the other hand, the second type of solutions can reach both very big maximum mass up to $ \sim 3.25 M_{\odot}$ and quite small radii $ R_{1.6}\lesssim 11.7$ km. The preferred solution, represented with a red dashed line in Fig. \ref{curves_2p_comparison}, belongs to the first class and thus the conformal limit on $c_s$ is fulfilled. We obtained for this EOS $M_{max}=2.13 M_{\odot}$ and $R_{1.6}= 12.20$ km. \\
Concerning the NN results, the optimal values are $e_0=191.29$ MeV fm$^{-3}$, $c_s^2 = 0.38$. Although this point do not exactly correspond to the absolute peak in the Bayesian Posterior, it is one of the multiple high probability modes of the distribution and it is located well inside the largest likelihood region.
The reason for the existence of many relevant combination of parameters, all belonging to the low ($e_0$, $c_s^2$) class, is that the $M-R$ sequences corresponding to each of them coincide for the most part of the sequences. This appears evident from the red and blue dashed lines of Fig. \ref{curves_2p_comparison} representing the preferred EOSs found using the two approaches. The NN curve is indeed characterized by a $R_{1.6}=12.26$ but with a larger maximum mass $M_{max}=2.37 M_{\odot}$. 
We underline that for a total mass as that of GW190425 and the preferred parameters found in these analysis, a prompt collapse is expected for a double QSs binary.
Similar to the $1P$ case, our results in the $M-R$ and $M-\Lambda$ planes are not in agreement with the 68\% CI of the GW170917\_1 PDF and only marginally compatible with the source J0437–4715. \\
After the training, The RMSEs of the NN found on the test set are  $15.5$ MeV fm$^{-3}$ for $e_0$  and $0.13$ for $c_s^2$. As mentioned before, the error is estimated as the 68\% CI of the 2D Gaussian built from the single RMSEs on the parameters. This approximation is the main limitation of our NN approach: the absence of an explicit probability distribution prevents the correlation between $e_0$ and $c_s^2$ to be seen. As a consequence, we cannot find the two types of solutions as provided by the Bayesian method and the 68\% CIs in the M-R and M-$\Lambda$ planes are considerably larger.

 \section{Summary and Conclusions} \label{summary}
In this work, we parameterized the QS EOS adopting a constant-speed-of-sound model with two parameters $(c_s^2, e_0)$. We used $M-R$ posteriors of several X-ray sources and $\Lambda$ posteriors from GW events to estimate the most probable values of those parameters using both Bayesian inference and NN prediction methods. For the NN calculations, we used marginalized Gaussians when incorporating the observational errors. While we have found that results from these two methods are in agreement with each other, from the construction of our NN we do not get any correlation between the predicted parameters. In contrast to the previous works of \cite{Fujimoto:2017cdo,Fujimoto:2019hxv}, we have tried to provide a quantitative comparison between these two methods given the QS EOS model. We have also included the tidal deformabilities at the same level of radius without converting it into other equivalent quantities while performing the NN predictions as done in \cite{Morawski:2020izm}. Both methods predict inconsistency of our EOS model with the sources GW170817\_1 and J0437--4715 at the 68\% level although other sources used in the studies are compatible with the predictions.  The compatibility found between the estimated parameter values from those two methods motivates to recognize NN based prediction as an efficient complimentary method to the standard Bayesian calculation. One of the criticism of this work can be the way we have incorporated the uncertainties of the measurements in the NN framework. Ideally, one would prefer to use the full distribution instead of a double Gaussian with marginalized data. That is one of our future plan to find out a computationally efficient procedure to include the comprehensive data sets. This will automatically include the correlation between the $M-R$ and $M-\Lambda$ measurements of the sources. \\ 
Moreover, we plan to improve our NN method to include a study of the correlations in our parameter space and therefore to obtain a better estimate of the errors.

\section*{Acknowledgments}
We would like to thank Giuseppe Pagliara and Alessandro Drago for useful discussions. P.C. acknowledges support from INFN postdoctoral fellowship. P.C. is supported by the Fonds de la Recherche Scientifique-FNRS, Belgium, under grant No. 4.4503.19.

\software{Keras \citep{chollet2015keras}, TensorFlow \citep{abadi2016tensorflow}, Scikit-learn \citep{pedregosa2011scikit}, corner.py \citep{corner}, emcee \citep{ForemanMackey:2012ig}}

\bibliography{mybiblio}

\end{document}